\documentclass[12pt,dvips]{article}
\usepackage{graphicx}
\begin{document}


\noindent {\bf Analysis of Stochastic Evolution}\\


\noindent{\small Francesco Vallone}\\


\noindent{\scriptsize Department of Economics, University of Turin, Via Po 53, Turin, Italy}\\


\begin{abstract}
Many studies in Economics and other disciplines have been reporting distributions following 
power-law behavior (i.e distributions of incomes (Pareto's
law), city sizes (Zipf's law), frequencies of words in long sequences of
text etc.)[1, 6, 7]. This widespread observed regularity has been explained in many ways: generalized Lotka-Volterra (GLV)  equations, self-organized criticality and highly optimized tolerance [2,3,4]. The evolution of the phenomena exhibiting power-law behavior is often considered to involve a varying, but size independent, proportional growth rate, which
mathematically can be modeled by geometric Brownian motion (GBM) $dX_t = r_t X_t dt + \alpha X_t W_t$
where $W_t$ is white noise or the increment of a Wiener process. It is the primary purpose of this article to study both the upper tail and lower tail of the distribution following the geometric Brownian motion and to correlate this study with recent results showing the emergence of power-law behavior from heterogeneous interacting agents [5]. The result is the explanation for the appearance of similar properties across a wide range of applications.      
\end{abstract}

\section{Introduction to power law distributions}
\noindent

This paper stems from the curiosity of the author to peer into the apparent similarities in the distribution of quite diverse phenomena, occurring in Social and Natural Sciences and usually explained by the use of skewed distributions. Scholars and scientists are used to measure variables pertinent to their fields, associating to each measure a particular {\it scale}. Often, as in the case of individual wealth across a population, the measurement can span several order of magnitude depending on the chosen individual. Similarly, cities size or firms size can present values distributed across multiple scales. In all these cases, the distribution representing the sample under investigation appears {\it right skewed}, i.e. while the majority of the measurements lie at lower level, there are quite a few measurements that are plotted well to the right that the typical value, creating a tail in the distribution. As a curiosity, we notice that in Economics this tail goes by the name of {\it fat tails}, in Physics it is referred to as {\it critical fluctuations}, in Computer Science and Biology it is {\it the edge of chaos}, and in Demographics and Linguistics it is called {\it Zipf's law}.

It has been reported in Economics since 1897 by Pareto that the distribution of personal income possesses, after a certain threshold, a heavy-tail following a power law (i.e. $\sim x^{-\alpha}$) [14]. In his work, Pareto indicated a value for the exponent $\alpha$ of about 1.5. The same behavior has been reported more recently in other disciplines. For example, the Harvard linguist Zipf discovered, in 1949, analyzing the frequencies of words in long sentences of text that they were distributed according to a power law with a value of the exponent $\alpha$ equal to one [7]. In this respect, Zipf's distribution (also known as zeta distribution) can be considered the discrete counterpart of the continuous Pareto's distribution.

Today, we are aware that many other phenomena have been reported following power laws with various values of the exponent $\alpha$ like clusters of Bose-Einstein condensate near absolute zero (following Pareto's law)[16], city sizes (following Zipf's law), the number of hits on a website, just to mention few cases [1,3, 6, 23]. 

Similar analysis for the distribution of income have showed in recent years a value of the exponent close to 1.4 (Pareto' law), for the case of Great Britain distribution of wealth [15], and to 1.0 (Zipf's law), for the case of American firm size as measured by receipts and employees [1]. A different result was obtained by Stanley et al., who reported that American firm size followed a lognormal distribution instead of a power law distribution [17].

Lognormal distributions were advanced, probably in the first real attempt to tackle this problem from a theoretical standpoint, by Gibrat in 1931. Gibrat first observed that assuming that the growth rate of each firm was independent of its size (Gibrat's law in weak form), the distribution of firms' size should have been right skewed and, if the rates of growth were only moderately correlated, such distribution should have been a member of the lognormal family (Gibrat's law in strong form) [18]. 

From an empirical point of view, it is known that growth rates of firms' output and countries GDP follow exponential densities or, as they are also known, Laplace distributions (i.e. $\sim \alpha \exp[-\alpha x], x>0$) [19,20]. This fact would confirm inherently Gibrat's law in weak form. The strong form of Gibrat's law is confirmed only by the study of Stanley et al. but is not confirmed by several other analysis starting from Pareto and ending, more recently, with Axtell. This paper aims at fixing these apparent inconsistencies of the theory with the associated empirical findings and at putting in perspective some triumphalism about natural laws.

\section{Explanations of power law distributions}
\noindent

An interesting quest for explaining the reasons of this apparent similarities is currently on-going and involves several scholars and researchers, all united in a multidisciplinary effort. This widespread observed regularity has been explained in many ways: Generalized Lotka-Volterra (GLV) equations, self-organized criticality (SOC) and highly optimized tolerance. 

GLV equations involve three scalar parameters and a probability distribution, each having
definite roles in the model's interpretation. These equations are suitable for multi-agents simulations. The first and second term of the equations are typically the auto-catalytic terms. They are, respectively, the growth rate of the system that is proportional to the size of the variable under investigation (i.e. individual wealth, city size etc.) and to the arithmetic mean of the same variable across the system, while the last term represents the mean-field interaction and is to be considered a competing term. The balance of the auto-catalytic terms with the competing term determines dynamical equilibrium in the system and power law distributions [2, 3]. 

If the system is far from equilibrium, self-organizing phenomena and a state of self-organized
criticality (SOC) may occur. According to the notion of SOC [21], scaling emerges because the sub-units of a system are heterogeneous and interact, and this leads to a critical state without any attractive points
or states. The occurrence of a power law may be read as a symptom of self-organizing
processes at work. A notable example of this approach applied to macroeconomics is the
inventory and production model developed by Bak et al. [22]. 

The classic example of highly optimized tolerance can be explained through a model of forest fires. Suppose that fires start at random in a grid-like forest and that, instead of appearing at random, trees are deliberately planted by a knowledgeable farmer. One can ask what the best distribution of trees is to optimize the amount of lumber the forest produces, subject to random fires that could start at any place. The answer turns out to be that one should plant trees in blocks, with narrow firebreaks between them to prevent fires from spreading. Moreover, one should make the blocks smaller in regions where fires start more often and larger where fires are rare. The reason for this is that we waste some valuable space by making firebreaks, space in which we could have planted more trees. If fires are rare, then on average it pays to put the breaks further apart so that more trees will burn if there is a fire, but we also get more lumber if the event do not occur. Carlson and Doyle show both by analytic arguments and by numerical simulations that, for quite general distributions of starting points for fires, this process leads to a distribution of fire sizes that approximately follows a
power law. The distribution is not a perfect power law, in this case. Carlson and Doyle have proposed that
highly optimized tolerance could be a model not only for forest fires but also for the sizes of files on the world wide
web, which appear to follow a power law [4, 24, 25]

Before starting our journey, in order to temper the enthusiasm behind power-law distributions, which are described too often as able to represent almost every complex phenomena, we would like to remind us the caveat clearly expressed by Prof. Feller commenting the good fit of the logistic distribution function with real life phenomena ``Lengthy tables, complete with chi-square tests, supported [the] thesis [that a logistic distribution function could be representative] for human populations, for bacterial colonies, development of railroads, etc. Both height and weight of plants and animals were found to follow the logistic law even though it is theoretically clear that these two variables cannot be subject to the same distribution. [...] The only trouble with the theory is that not only the logistic distribution but also the normal, the Cauchy and other distributions can be fitted to the {\it same material with the same or better goodness of fit}.''[8,9]

\section{The stochastic differential equation GBM}
\noindent

Since the phenomena we are going to investigate have a varying, but size independent, proportional growth rate, they are described by the following stochastic differential equation (GBM)
\begin{equation}
\frac{dX_t}{dt} = a_t X_t
\end{equation}
where $a_t = r_t + \alpha W_t$ so that $r_t=r$ represents the constant growth rate to which we have added white noise disturbance $W_t$ or the increment of a Wiener process. As we will see, this stochastic equation does not allow for power law distributions but only for lognormal distribution unless we revisit our understanding of the concept of time. 

The way $W_t$ is typically constructed is as a probability measure on the space of tempered distributions on $[0,\infty)$ and not as a probability measure on the much smaller space $ R^{[0,\infty)}$, like an ordinary process can. $W_t$ has {\it stationary independent increments} and is equivalent to a Brownian motion $B_t$ following a normal distribution with mean $0$ and variance $t$ [10,11,12,13].

Let us make use of the It{\^o} calculus and formally replace $W_t$ by $dB_t dt$ in equation [1]. We then obtain

\begin{equation}
dX_t = r_t X_t dt + \alpha X_t dB_t.
\end{equation}
The solution to this equation is 

\begin{equation}
\int_0 ^t \frac{dX_t}{X_t} = r_t t + \alpha dB_t,
\end{equation}
where we have assumed $B_0=0$.

The integral on the left side must be evaluated using the It{\^o}'s formula for stochastic integrals 
so that 
\begin{equation}
d \ln X_t = \frac1{X_t} dX_t + \frac12 (-\frac{(dX_t)^2}{X_t^2})= \frac{dX_t}{X_t}-\frac12\alpha^2 dt
\end{equation}
and, after a final integration,
\begin{equation}
\ln X_t = \ln X_0+ (r-\frac12\alpha^2)t + \alpha B_t.
\end{equation}

Equation (5) is the functional form that links the two variables $X_t$ and $B_t$. We can easily proceed on deriving the distribution of $\ln(X_t)$ from that one of $B_t$, after a certain threshold value $T$. 

It is a trite calculation to obtain 
\begin{equation}
\ln X_T \rightarrow N \left[ \ln X_0+ (r-\frac12\alpha^2)T, \alpha^2 T\right],
\end{equation}
where $N$ represents the normal distribution. We have then proved that GBM can provide only lognormal distributions.

Let us assume in a way similar to what has been presented by Reed [26] that the threshold value $T$ is a random variable itself, distributed according to an exponential law or Laplace law, $f_T(t)=\nu \exp[-\nu t], t>0$. In a certain sense, it is like saying that the evolution of the system (to fix the ideas let us take firms size) follows the GBM for {\it an interval of time}, which is itself a random variable and can be related to the {\it time} of observation. 

With this assumption, we have that the moments generating function of the variable $\ln X_T$ and $f_T$ are respectively
\begin{equation}
E[e^{\ln X_T(t) s}] = \exp\left\{(\ln X_0 s) + \left[(r-\frac12\alpha^2)s+ \frac12\alpha^2 s^2\right]T\right\}
\end{equation}

and

\begin{equation}
E[e^{f_T(t) s}] = \frac{\nu}{\nu - s},
\end{equation}

where $E$ represents the functional of the mean value operator.

It is an easy derivation, following Reed [26], to find that the state after time $T$, indicated by a tilde, $\ln\tilde{X}_T$ is no longer lognormally distributed.

In fact, we have

\begin{equation}
E_T[E(e^{\ln \tilde{X}_T s} | T)] =\frac{\nu e^{\ln \tilde{X}_0 s}}{\left\{\nu + \left[(r-\frac12\alpha^2)s+ \frac12\alpha^2 s^2\right]\right\}}
\end{equation}

Let us denote the two positive solutions $m_1$ and $-m_2$ of the quadratic equation
$\frac12\alpha^2 s^2 -(r-\frac12\alpha^2)s-\nu = 0$ so that we can write

\begin{equation}
E_T[E(e^{\ln \tilde{X}_T s} | T)] = e^{\ln \tilde{X}_0 s} \frac{m_1 m_2}{(m_1 -s) (m_2 + s)}
\end{equation}

It can be verified that the former equation is the moments generating function of the asymmetric Laplace
distribution centered on $\ln \tilde{X}_0$ with probability density function
\begin{eqnarray}
	f_{\ln \tilde{X}_0}(\xi)=\frac{m_1 m_2}{m_1 + m_2} e^{m_2(\xi - \ln \tilde{X}_0)}, \xi\le \ln \tilde{X}_0\\
	f_{\ln \tilde{X}_0}(\xi)=\frac{m_1 m_2}{m_1 + m_2} e^{-m_1(\xi - \ln \tilde{X}_0)}, \xi\ge \ln \tilde{X}_0
\end{eqnarray}

From the previous distributions, we can deduce that $X_0$ has the following distribution functions
\begin{eqnarray}
	f_{\tilde{X}_0}(x)=\frac{m_1 m_2}{m_1 + m_2} \left(\frac{x}{\tilde{X}_0}\right)^{m_2-1}, x\le \tilde{X}_0\\
	f_{\tilde{X}_0}(x)=\frac{m_1 m_2}{m_1 + m_2} \left(\frac{x}{\tilde{X}_0}\right)^{-m_1-1}, x\ge \tilde{X}_0
\end{eqnarray}
showing power law behavior in both tails (Reed refers to this distribution by the expression {\it double-Pareto distribution} [26]).

\section{Analysis of the double-Pareto distribution} 
\noindent

An original part of our paper is represented by the exploration of the possible values of $m_1$ and $m_2$ and of their determinants.

Solving the quadratic equation $\frac12\alpha^2 s^2 -(r-\frac12\alpha^2)s-\nu = 0$, we obtain

\begin{eqnarray}
	m_1= \frac{(r-\frac{\alpha^2}{2})}{\alpha^2} \left( 1+ \sqrt{1+ 8 \frac{\nu \alpha^2}{(2 r-\alpha^2)^2}}\right)
\end{eqnarray}
\begin{eqnarray}
	m_2= \frac{(r-\frac{\alpha^2}{2})}{\alpha^2} \left(1- \sqrt{1+ 8 \frac{\nu \alpha^2}{(2 r-\alpha^2)^2}}\right) 
\end{eqnarray}

Let us start considering the limits
\begin{eqnarray}
	\lim_{\nu\rightarrow 0} m_1= \frac{(2 r - \alpha ^2)}{\alpha^2}\\
	\lim_{\nu\rightarrow 0} m_2= 0
\end{eqnarray}

and

\begin{eqnarray}
	\lim_{\nu\rightarrow \infty} m_1= \infty\\
	\lim_{\nu\rightarrow \infty} m_2= -\infty
\end{eqnarray}

Few considerations are in order. First of all, the case in which $\nu$ gets smaller means that the time between two consecutive evolutions according to the GBM model gets longer and, therefore, the phenomenon turns to be exclusively described by a lognormal distribution. In fact, it is more correct to affirm that the distribution is no more a power law but rather a lognormal distribution, even if we save only one of the two solutions that can make sense for certain $r$ and $\alpha$. When $\nu$ gets larger, the model explodes, providing no further predicting power.

Let us continue estimating the following limits that will provide enough information to plot these functions in terms of $\alpha$ (see figure 1)

\begin{eqnarray}
	\lim_{\alpha\rightarrow 0} m_1= \infty,\,provided\,\, r>0\\
	\lim_{\alpha\rightarrow 0} m_2= -\frac{\nu}{r}
\end{eqnarray}

and

\begin{eqnarray}
	\lim_{\alpha\rightarrow \sqrt{2r}^+} m_1= +\sqrt{\frac{\nu}{r}}\\
	\lim_{\alpha\rightarrow \sqrt{2r}^-} m_1= -\sqrt{\frac{\nu}{r}}\\
	\lim_{\alpha\rightarrow \sqrt{2r}^+} m_2= -\sqrt{\frac{\nu}{r}}\\
	\lim_{\alpha\rightarrow \sqrt{2r}^-} m_2= +\sqrt{\frac{\nu}{r}}
\end{eqnarray}

and, finally,
	
\begin{eqnarray}
	\lim_{\alpha\rightarrow \infty} m_1= -1\\
	\lim_{\alpha\rightarrow \infty} m_2= 0
\end{eqnarray}

It is quite interesting to see that $\alpha_{\star} = \sqrt{2 r}$ separates two kinds of solutions: to the left of this point $m_1$ is positive and $m_2$ is negative, while to the right of this point the opposite occurs. Considering the particular symmetry of the solutions, we can interchange $m_1$ with $m_2$. We will say, then, that $\alpha_{\star}$ is defined as {\it the natural degree of stochasticity} of the system and is proportional to the square root of the growth rate of the system.  

Another point that deserves our attention is the role played by the ratio between $\nu$ and $r$, which are both the inverse of time. Essentially, these parameters measure respectively the time scale of the sampling of our system and of its growth rate. Based upon real life considerations, we can assume that the latter one is much greater than the former one and, therefore, put $\nu/ r<< 1$.

This results can be translated into two separate regimes:
\begin{enumerate}
	\item for $\alpha < \alpha_{\star}$
		\begin{equation}
			m_1\in (\sqrt{\nu/r}\approx 0, \infty), m_2\in(-\sqrt{\nu/r}\approx 0,-\nu/r= 0)
		\end{equation} 
	\item for $\alpha > \alpha_{\star}$
		\begin{equation}
			m_1\in (-\sqrt{\nu/r}\approx 0, -1), m_2\in(\sqrt{\nu/r}\approx 0,0)
		\end{equation} 
\end{enumerate}

Considering that $\alpha_{\star}$ can be associated to the natural degree of stochasticity of the system that is proportional to the square root of the growth rate of the system, while $\alpha$ can be referred to the {\it internal degree of stochasticity} of the system, we will then separate between two kinds of system: {\it quasi-stochastic} systems, defined by equation (29), and {\it stochastic} systems, defined by equation (30).

\noindent

\begin{figure}
	\caption{Plot of $m_1$ and $m_2$ as a function of $\alpha$.}
	\label{fig:m1m2}
\end{figure}

It is definitely clear that quasi-stochastic systems have $m_1$ that can assume almost any positive value, while $m_2\approx 0$, that is these systems might have one of the two tails following a Zipf law. 

On the other side, stochastic systems are characterized by $m_1$ that results confined between $0$ and $-1$ and by $m_2\approx 0$. This is the realm where most of the previous reported examples live.

At this point, there should be no surprise in the fact that empirical data show both Pareto's and Zipf's laws because the exact exponent depends on the ratio between the frequency of the sampling and the growth rate of the system, which depends ultimately on the degree of inherent stochasticity of the particular system considered, which is proportional to the growth rate of the system.

\section{Degree of Stochasticity and Heterogeneous Interacting Agents (HIA)}
\noindent

We have seen in the previous section the existence of an internal degree of stochasticity associated with the parameter $\alpha$. We can talk of a stochastic system whenever $\alpha$ results greater than the natural degree of stochasticity of the system defined by $\alpha_{\star}=\sqrt{2r}$.

Most of the systems considered in the examples of Section 1 are made up of agents. Agents are typically considered alike and non-interacting but in doing so there is a considerable lack of realism. Thanks to the new venue of agent-based simulation model, we can develop agents that are different and interacting. Agents are processes implemented on a computer that have autonomy (they control their own actions); social ability (they interact with other agents through some kind of 'language'); reactivity (they can perceive their environment and respond to it); and pro-activity (they are able to undertake goal-directed actions) [27]. 

Starting from an agent-based model, it is possible to achieve a micro-foundation of the system and, as a matter of facts, of its degree of stochasticity that will depend among other things by the number of agents and by the quality and features of the interactions among the agents. In all these systems, we would expect, having from one side an operational GBM system and from the other an observer, sampling key parameters with the consequent appearance of power laws.

For this reason, it was no surprise for us to read that Delli Gatti et al. reported that using agents-based simulation model in which agents were affected by financial fragility observed as a consequence of business fluctuation the appearance of power law [5]. Similar universalities have been found by the author in an agent-based simulation model in which firms interact with other firms adjusting the prices and quantities of their products in order to survive (i.e. achieve sales in excess of total costs)[28]. In it, the firms sizes are distributed according a power law. Changing the degree of internal stochasticity of the simulation, by changing the level of interaction of the agents, the number of the agents and/or the intensity of the interaction, alters the exponent of the distribution as predicted by the model presented in the previous section.

\section{Conclusions}
\noindent

Complex systems evolve according to stochastic models that can simply be modeled by a GBM equation. Together with the evolution of the system there is always a definite time of observation and evolution itself that must be considered a random variable as well. This fact changes drastically the natural distribution function of the GBM model (i.e. a lognormal distribution) providing a power law at its place. 

The exponent of the derived power distributions has several features depending both on the comparative level of internal degree of stochasticity of the system versus its natural degree that is proportional to the square of the growth rate and on the ratio between the frequency of the evolution of the system according to GBM and its growth rate. Thanks to these key parameters, we constructed a considerable versatile model, capable of great consistency between theory and empirical data and with strong predicting power.

The final link between theory and empirical data is stressed by recent contributions provided by agents-based simulation models because they can generate in a ``lab-like environment'' stochastic evolutionary systems and micro-foundation. 

Our journey has not terminated with this paper, although we have provided a small contribution to a deeper understanding of systems exhibiting stochastic evolution. These systems are all around us and represent the complexity of life. Economic systems, which are the object of our present research, have all the characteristics highlighted in the article and show some of the evidence found in previous equations. Even if we are yet totally unconvinced that scholars should refer to these regularities among diverse phenomena as universal laws, we have showed that it might exist a common pattern around which several and diverse phenomena can find an easy explanation.

In order to continue the present effort, we will need to develop a better understanding of the correlation between natural and internal degrees of stochasticity of the system and its fundamental parts. The presence of heterogeneous interacting agents seem to be a powerful model to capture the complexity of stochastic systems.

\section{References}

\vspace{6mm}

{\baselineskip 5mm\footnotesize
\noindent [1] R. Axtell, {\it Zipf distribution of U.S. firm sizes} in Science {\bf 293} (2001), 1818-1820

\noindent[2] S. Solomon, {\it Generalized Lotka-Volterra Equation} in Econophysics {\bf 97} (1998), eds. Imre Kondor and Janos Kertes

\noindent [3] X. Gabaix, {\it Zipf's law for cities: an explanation} in Quarterly Journal of Economics {\bf 114} (1999), 739-767

\noindent [4] N. Newman, {\it The power of design} in Nature {\bf 405} (2000), 412-413

\noindent [5] D. Delli Gatti, C. Di Guilmi, E. Gaffeo, G. Giulioni, M. Gallegati, A. Palestrini {\it A new approach to business fluctuations: heterogeneous interacting agents, scaling laws and financial fragility} in Journal of Economic Behavior and Organization {\bf 56} (2005), 489-512

\noindent [6] R. Kali {\it The city as a giant component: a random graph approach to Zipf's law} in Applied Economics Letters, 15 September (2003), {\bf 10}, Iss. 11, 717-720 

\noindent [7]	K. George, Zipf, Human Behaviour and the Principle of Least-Effort, Addison-Wesley, Cambridge MA, 1949 

\noindent [8] W. Feller, An introduction to Probability Theory and Its Applications, John Wiley and Sons, Vol. 2, II Ed. 1971

\noindent [9] W. Feller, {\it On the logistic law of growth and its empirical verifications in biology}, in Acta Biotheorica, {\bf 5} (1940), 51-66

\noindent [10] T. Hida, Brownian Motion, Springer-Verlag (1980)

\noindent [11] R.J. Adler, The Geometry of random fields,Wiley and Sons, (1981)

\noindent [12] Yu.A. Rozanov, Markov Random Fields, Springer-Verlag, (1982)

\noindent [13] T. Hida, H.H. Kuo, J. Potthof and L. Streit, White Noise: an infinite dimensional approach, Kluvert, (1993)

\noindent [14] V. Pareto, Course d'Economie Politique, vol. 2. Pichou, Paris, (1897)

\noindent [15] S. Solomon, M. Levy, {\it Power laws are logarithmic Boltzmann laws}, J. Mod. Phys. C, {\bf 7}, p. 745 (1996)

\noindent [16] M.S. Watanabe, Phys. Rev. E, {\bf 53}, 4187 (1996)

\noindent [17] M.H. R. Stanley et al., Econ. Lett., {\bf 49}, 453 (1995)

\noindent [18] R. Gibrat, {\it Les In{\'e}galit{\'e}s Economiques, Applications aux In{\'e}galit{\'e}s des Richesses, {\'a} la Concentration des Entreprises, aux Populations des Ville, aux Statistiques des Familles, etc., d'une Loi Nouvelle, la Loi de l'Effet Proportionnel} Librarie du Recueil Sirey, Paris, (1931)

\noindent [19] M.H. Stanley, L. Amaral,S. Buldyrev, S. Havlin, H. Leschorn, P. Maas, M. Salinger,E. Stanley, {\it Scaling behavior in the growth of companies} Nature {\bf 379}, 804–806 (1996)

\noindent [20] L. Amaral, S. Buldyrev, S. Havlin, H. Leschhorn, P. Maas, M. Salinger, E. Stanley, M. Stanley,{\it Scaling behavior in Economics: I. Empirical results for company growth.}, Journal de Physique {\bf 7}, 621–633 (1997)

\noindent [21] P. Bak, How Nature Works. Oxford University Press, Oxford (1997)

\noindent [22] P. Bak, K. Chen, J. Scheinkman, M. Woodford, {\it Aggregate fluctuations from independent sectoral
shocks: self-organized criticality in a model of production and inventory dynamics.} Ricerche Economiche {\bf 47},
3–30 (1993)

\noindent [23] L. A. Adamic and B. A. Huberman, {\it The nature of markets in the World Wide Web.}, 
Quarterly Journal of Electronic Commerce {\bf 1}, 512 (2000)

\noindent [24] M. E. J. Newman, {\it Power laws, Pareto distributions and Zipf's law.}, Contemporary Physics 46, 323-351 (2005)

\noindent [25] J. M. Carlson and J. Doyle, {\it Highly optimized tolerance:A mechanism for power laws in designed systems.}, Phys. Rev. E {\bf 60}, 1412–1427 (1999). 

\noindent [26] W. Reed,  {\it The Pareto, Zipf and other power laws.} Economics Letters {\bf 74}, 15–19. (2001)

\noindent [27] M. Wooldridge and N.R. Jennings, {\it Intelligent agents: theory and practice}, Knowledge Engineering Review, Vol.10, pp.115-152 (1995)

\noindent [28] F. Vallone, {\it Knowledge Economics: experimental price and production adjustments allow survival of firms}, Working Paper, (2005) 

\end{document}